\documentclass[runningheads]{llncs}
\usepackage[T1]{fontenc}
\usepackage{amsmath}
\usepackage{amsfonts}
\usepackage{amssymb}
\usepackage{graphicx,hyperref}
\usepackage{bussproofs}
\usepackage{mathrsfs,multicol}
\usepackage{enumerate,longtable}

\newcommand{\M}{\mathfrak{M}}

\newcommand{\RL}{{\bf RL}}
\newcommand{\RLL}{{\bf RL^2}}
\newcommand{\VAR}{\mathit{VAR}}
\newcommand{\PAR}{\mathit{PAR}}
\newcommand{\CON}{\mathit{CON}}
\newcommand{\Lii}{\mathscr{L}}
\newcommand{\Fii}{\mathscr{F}}
\newcommand{\Li}{\mathcal{L}}
\newcommand{\Fi}{\mathcal{F}}
\begin{document}
\title{On a Second-Order Version of Russellian Theory of Definite Descriptions}
\author{Yaroslav Petrukhin\inst{1}\orcidID{0000-0002-7731-1339}}
\authorrunning{Y. Petrukhin}
\institute{University of \L{}\'{o}d\'{z}, \L{}\'{o}d\'{z}, Lindleya 3/5, 90-131,  Poland \\
\email{yaroslav.petrukhin@gmail.com}}
\maketitle              % typeset the header of the contribution
\begin{abstract}
Definite descriptions are first-order expressions that denote unique objects. In this paper, we propose a second-order counterpart, designed to refer to unique relations between objects. We investigate this notion within the framework of Russell's theory of definite descriptions. While full second-order logic is incomplete, its fragment defined by Henkin's general models admits completeness. We develop our theory within this fragment and formalize it using a cut-free sequent calculus.

\keywords{Definite descriptions  \and Second order logic \and Proof theory
\and Lambda-abstraction \and Sequent calculi \and Cut admissibility}
\end{abstract}
\section{Introduction}
Definite descriptions are usually expressed as first-order constructs in the form $ \iota x \varphi $, where $ \iota $ is a term-forming operator, $ x $ is an individual variable, and $ \varphi $ is a formula. The core idea behind definite descriptions is to denote the unique object $ x $ that satisfies the formula $ \varphi $. In this work, we propose a second-order extension of this concept: $ \iota X \varphi $, where $ X $ is a relational variable. Such expressions are meant to signify the unique \textit{relation} for which the formula $ \varphi $ holds true.

First-order quantifiers deal with objects, whereas second-order quantifiers concern the properties of objects or relations between objects. We utilize this relationship between first-order and second-order quantifiers when introducing second-order definite descriptions. It is noteworthy that prior efforts have been made to examine definite descriptions in the realm of second- or higher-order logics, as conducted by Makarenko and Benzm\"{u}ller \cite{MakarenkoBenzmuller}. Nonetheless, in their analysis, definite descriptions continue to function as first-order expressions pertaining to objects. 

Let us present several examples of the second-order definite descriptions. The first example is the transitive closure of a graph:
$\iota R\; \mathsf{TransitiveClosure}(R, G)$. 
    This describes the unique binary relation $R$ that forms the transitive closure of the edge relation in graph $G$. The expression $$\iota P\; \left( \mathsf{Path}(P, a, b, G) \land \forall P'\left(\mathsf{Path}(P', a, b, G) \rightarrow \mathsf{Length}(P) \leq \mathsf{Length}(P')\right) \right)$$ represents  `the shortest path relation between two nodes $ a $ and $ b $ in graph $ G $', where $P$ is a path (represented as a relation), and the description denotes the unique shortest path from node $a$ to node $b$ in graph $G$. The connectivity relation in a graph can be formalized as $ \iota R\; \left( \forall x\forall y\; \left( R(x,y) \leftrightarrow \exists P\; \mathsf{Path}(P, x, y, G) \right) \right) $. Finally,  the expression $ \iota R\; \left( \mathsf{TotalOrder}(R) \land \forall x\forall y\; \left( P(x,y) \rightarrow R(x,y) \right) \right) $ denotes the total order extending a given partial order 
    $ P $, assuming such an extension is unique.

Russellian theory of definite descriptions \cite{Russell,WhiteheadRussell} is arguably one of the most recognized and frequently accepted, notwithstanding the criticism it has received. It possesses several disadvantages; nevertheless, there are methods to mitigate at least some of them. 
We follow the presentation of Russellian theory as articulated by Indrzejczak and Zawidzki \cite{IndrzejczakZawidzkiRL} and Indrzejczak and K\"{u}rbis \cite{IKRL}. In Russellian theory,  definite descriptions are characterized by the following formula, where $\psi$ must be limited to atomic formulas, unless additional mechanisms are introduced to mark scope distinctions:
\[
 \psi(\iota y\varphi) 
% \psi[x/\iota y\varphi]
 \leftrightarrow \exists x (\forall y (\varphi \leftrightarrow y = x) \land \psi)
\]

Indrzejczak, Zawidzki, and K\"{u}rbis' approach characterize them with the help of $\lambda$-operator as follows: 
\[
 (\lambda x \psi)\iota  y \varphi \leftrightarrow \exists x (\forall y (\varphi \leftrightarrow y = x) \land \psi)
\]

This approach allows both complex and primitive predicates to be applied to definite descriptions while avoiding scope-related issues. In particular, it helps us answer the question: in the negated expression $ \neg \psi(\iota y, \varphi) $, does the negation apply to the entire expression, or solely to $ \psi $? Whitehead and Russell, seeing the issue, proposed the method of scope distinctions \cite{WhiteheadRussell}. Nonetheless, it is recognized for its clumsiness. One may attempt to circumvent the issue by limiting the formulas to atomic ones; however, this considerably diminishes the expressive capacity of Russellian theory. Nonetheless, this methodology has been employed by Kalish, Montague, and Mar \cite{KalishMontagueMar} as well as Francez and Wi\k{e}ckowski \cite{FrancezWieckowski} in their natural deduction systems, and Indrzejczak \cite{IndrzejczakRussel} in his cut-free sequent calculus. 
 Another concept has been proposed by K\"{u}rbis \cite{KurbisBQ}. 
He implements a binary quantifier represented as $ I x[\varphi,\psi] $.
This resolves the issues; but, if one wants to consider definite descriptions as terms, an alternative approach is required. This inspired Indrzejczak, Zawidzki, and K\"{u}rbis \cite{IndrzejczakZawidzkiRL,IKRL} to apply predicate abstracts of the form $\lambda x\varphi$ (`the property of being $ \varphi $') to terms, including definite descriptions, to obtain formulas called lambda atoms. They observe that predicate abstracts built by means of the lambda-operator have been previously utilized by Stalnaker and Thomason \cite{StalnakerThomason}, Bressan \cite{Bressan}, Fitting \cite{Fitting}, Scales \cite{Scales}, Fitting and Mendelsohn \cite{FittingMendelsohn}, and Indrzejczak \cite{Indrzejczak2020}.

Another source for motivation for the introduction of $\lambda$ is that if $\psi$ is complex in the Russelian formula characterizing definite descriptions, one may readily encounter a contradiction. The application of $\lambda$ addresses this problem, while a similar outcome could be attained through the adoption of free logic (rendering the entire theory deductively weaker and incapable of inferring contradictions; see, e.g. \cite{IndrzejczakFree})  or the utilization of paraconsistent logic \cite{PetrukhinNelsonDD}. 

This study adheres to the framework established by 
Indrzejczak and K\"{u}rbis \cite{IKRL} in presenting the semantics and employs their cut-free sequent calculus, which is adequate with respect to this semantics. We extend their approach by incorporating second-order quantifiers, a second-order variant of identity, and ultimately a second-order interpretation of lambda terms and definite descriptions. Unlike them, we do not provide a constructive proof of the cut admissibility theorem, as we believe this subject, due to its complexity, requires a separate paper.\footnote{The cut admissibility theorem for second- and higher-order logics has remained an unresolved issue in proof theory for an extended duration, referred to as Takeuti's conjecture \cite{Takeuti1953}. Various scholars, utilizing distinct methodologies, reached a positive resolution: Tait \cite{Tait1966}, Prawitz \cite{Prawitz1968},  Takahashi \cite{Takahashi1967}, Girard \cite{Girard1971}. See also Takeuti \cite{Takeuti1975} and Rathjen and Sieg \cite{RathjenSieg} on this issue. Developing a syntactic constructive proof remains an open problem.} 
However, we provide a semantic proof of this statement obtained as a consequence of a Hintikka-style completeness proof in the spirit of \cite{AvronLahav,LahavAvron}. 

The structure of the paper is as follows. Section~\ref{Lang} introduces the languages and semantics of the logics under consideration. In Section~\ref{ND}, we present the corresponding sequent calculi and prove soundness, completeness, and cut admissibility. Section~\ref{Concl} concludes the paper.

\section{Languages and Semantics}\label{Lang}
We start with a description of the logic $\RL$ from \cite{IndrzejczakZawidzkiRL,IKRL}, which corresponds to the formulation of the Russellian theory involving $\lambda$ discussed above. 

\textit{The language $\Li$ of} $\RL$ is a standard first-order language with identity and without function symbols, but supplied with $\lambda$ and $\iota$. 
The language is built from two disjoint sets of symbols: $\VAR$, representing variables, and $\PAR$, representing parameters. In the proof-theoretic framework of $\RL$, elements of $\VAR$ are used exclusively as bound variables, while $\PAR$ provides the symbols for free variables. The language contains the set $\CON$ of constant symbols as well. In contrast, the semantic framework does not enforce this distinction. The basic terms of the language consist of variables, parameters, and constants. Additionally, we allow expressions formed using the definite description operator $\iota$ applied to predicate abstracts. These are referred to as quasi-terms. ``We mention only the following formation rules for the more general notion of a formula used in the semantics'' \cite[p. 115]{IKRL}:
\begin{itemize}
    \item If $P^n$ is an $ n $-ary predicate symbol (including $=$) and $t_1, \ldots, t_n \in \VAR \cup \PAR\cup \CON$, then $P^n(t_1, \ldots, t_n)$ is a formula (atomic formula).
    \item If $\varphi$ is a formula and $ x\in\VAR $, then $(\lambda x \varphi)$ is a predicate abstract.
    \item If $\varphi$ is a formula  and $ x\in\VAR $, then $\iota x\varphi$ is a quasi-term.
    \item If $\varphi$ is a predicate abstract and $t$ a term or quasi-term, then $\varphi t$ is a formula (lambda atom).
\end{itemize}

We write $\Fi$ for the set of all formulas of $\Li$, $x,y,z,x_1,\ldots$ for the members of $\VAR$, $a,b,c,a_1,\ldots$ for the elements of $\PAR$, \( k, k_1, k_2, \dots \) for the elements of $\CON$, $\varphi_t^x$ for the result of replacing $x$ by $t$ in $\varphi$, and, similarly, $\varphi_{t_1,\ldots,t_n}^{x_1,\ldots,x_n}$ for the result of a simultaneous replacing $x_1,\ldots,x_n$ by $t_1,\ldots,t_n$ in $\varphi$. When $t$ is a variable $y$, we assume that $y$ is free for $x$ in $\varphi$, meaning that the substitution does not cause any formerly free occurrence of $y$ to become bound within $\varphi$. We presume a similar condition for $\varphi_{t_1,\ldots,t_n}^{x_1,\ldots,x_n}$.

\textit{The semantics of} $\RL$. We define the notion of a model, following \cite[p. 115f]{IKRL}. A \textit{model} is a structure $M = \langle D, I \rangle$, where for each $n$-argument predicate $P^n$, $I(P^n) \subseteq D^n$. An \textit{assignment} $v$ is a function $v : \VAR \cup \PAR \to D$. An \textit{$x$-variant} $v'$ of $v$ agrees with $v$ on all arguments, save possibly $x$. We write $v_o^x$ to denote the $x$-variant of $v$ with $v_o^x(x) = o$. The notion of \textit{satisfaction} of a formula $\varphi$ \textit{with} $v$, in symbols $M, v \models \varphi$, is defined as follows, where $t \in \VAR \cup \PAR\cup \CON$:
\[
\begin{aligned}
M, v \models P^n(t_1, \ldots, t_n) & \text{ iff } \langle v(t_1), \ldots, v(t_n) \rangle \in I(P^n) \\
M, v \models t_1 = t_2 & \text{ iff } v(t_1) = v(t_2) \\
M, v \models (\lambda x \psi)t & \text{ iff } M, v_o^x \models \psi, \text{ where } o = v(t) \\
M, v \models (\lambda x \psi)\iota y\varphi & \text{ iff } \text{there is an } o \in D \text{ such that } M, v_o^x \models \psi, \\
& \quad M, v_o^x \models \varphi^y_x, \text{ and for any } y\text{-variant } v' \text{ of } v_o^x, \\
& \quad \text{if } M, v' \models \varphi, \text{ then } v'(y) = o \\
M, v \models \neg \varphi & \text{ iff } M, v \not\models \varphi \\
M, v \models \varphi \land \psi & \text{ iff } M, v \models \varphi \text{ and } M, v \models \psi \\
M, v \models \forall x \varphi & \text{ iff } M, v_o^x \models \varphi, \text{ for all } o \in D. 
\end{aligned}
\]

The truth conditions for $\vee$, $\to$, $\leftrightarrow$, and $\exists$ are standard. As for the constants, we postulate that $I(k)\in D$, for each constant $k$. The notions of satisfiable and valid formulas are defined in a standard way. The consequence relation is understood as follows, for all $\Gamma\subseteq\Fi$ and $A\in\Fi$: $\Gamma\models_\RL \varphi$ iff in every model $M$ and every assignment $v$, if $M,v\models \psi$, for all $\psi\in\Gamma$, then $M,v\models \varphi$. 

Let us now describe the logic $\RLL$, a second-order generalization of $\RL$.

\textit{The language} $\Lii$ \textit{of} $\RLL$ is a second-order extension of $\Li$. In addition to individual variables and parameters, we have the sets $\VAR^2=\{X,Y,Z,X_1,\ldots \}$ and $\PAR^2=\{A,B,C,A_1,\ldots \}$ of $n$-ary relational variables and parameters,   respectively (unary ones might be called property variables and parameters).  As in the first-order case, this distinction is important for proof theory, but might be relaxed in the case of semantics. In addition to individual constants, we have the set $\CON^2=\{K, K_1, K_2,\ldots \}$ of relational constants. They represent fixed \( n \)-ary relations over individual constants. The terms are constants and individual variables/parameters. Notice that relational variables/parameter are not terms. Atomic formulas are as follows: $t_1 = t_2$, $P(t_1, \ldots, t_n)$, $ X=Y $, and $X(t_1, \ldots, t_n)$, if $ t_1, \ldots, t_n $ are terms, $P$ is an $n$-ary relation symbol, and $X$ and $ Y $ are $n$-ary relational variables/parameters.\footnote{The formula $ X=Y $ is understood as $\forall x_1 \ldots \forall x_n \big(X(x_1, \ldots, x_n) \leftrightarrow Y(x_1, \ldots, x_n)\big)$. The formula $t_1 = t_2$ might be defined as $\forall X(X(t_1)\leftrightarrow X(t_2))$.} In addition to the above described atomic and first-order formulas, we define the following ones:
\begin{itemize}
\item If $\varphi$ is a formula and $X\in \VAR^2$, then $\forall X\varphi$ and $\exists X\varphi$ are formulas.
    \item If $\varphi$ is a formula  and $X\in \VAR^2$, then $(\lambda X \varphi)$ is a \textit{relational} abstract.
    \item If $\varphi$ is a formula  and $X\in \VAR^2$, then $\iota X\varphi$ is a \textit{pseudo}-term.
    \item If $\varphi$ is a relational abstract and $t$ is a  pseudo-term, then $\varphi t$ is a formula.
\end{itemize}

We write $\Fii$ for the set of all formulas of $\Lii$,  $\varphi_P^X$ for the result of replacing $X$ by a predicate symbol $P$ in $\varphi$. 

\textit{The first version of the semantics} (\textit{without a complete calculus}). In a model $M = \langle D, I \rangle$, an assignment $v$ should be redefined as follows\footnote{Formally, this makes $v$ a function with a dependent domain.}: 
$v(x)\in D$, for $x\in\VAR\cup\PAR$, and $v(X)\subseteq D^n$, for $X\in\VAR^2\cup\PAR^2$,
 an \textit{$X$-variant} $v'$ of $v$ agrees with $v$ on all arguments, save possibly $X$. We write $v_O^X$ to denote the $X$-variant of $v$ with $v_O^X(X) = O$, where $O\subseteq D^n$. The definition of the notion of satisfaction of a formula $\varphi$ with $v$ is extended by the following cases, where $t \in \VAR \cup \PAR$:
\[
\begin{aligned}
M, v \models X(t_1, \ldots, t_n) & \text{ iff } \langle v(t_1), \ldots, v(t_n) \rangle \in v(X),  \text{ if } X \text{ is } n\text{-ary},\\
M, v \models X=Y & \text{ iff }
v(X)=v(Y),\\
M, v \models (\lambda X \psi)\iota Y\varphi & \text{ iff } \text{there is an } O \subseteq D^n \text{ such that } M, v_O^X \models \psi, \\
& \quad M, v_O^X \models \varphi^Y_X, \text{ and for any } Y\text{-variant } v' \text{ of } v_O^X, \\
& \quad \text{if } M, v' \models \varphi, \text{ then } v'(Y) = O \\
M, v \models \forall X \varphi & \text{ iff } M, v_O^X \models \varphi, \text{ for all } O \subseteq D^n,\\
M, v \models \exists X \varphi & \text{ iff } M, v_O^X \models \varphi, \text{ for some } O \subseteq D^n.
\end{aligned}
\]

This semantics lacks the completeness theorem. In order to obtain this theorem, we need to deal with a fragment of the second-order logic: we should restrict the interpretations of the relational variables/parameters. Henkin's \cite{Henkin1950} concept of a general model will help us with this issue.

\textit{The second version of the semantics} (\textit{with a complete calculus}). \textit{The semantics of $\RLL$}. Although second-order logic is known to be incomplete, there exists a fragment that is complete with respect to general models \cite{Henkin1950} (see \cite{Stanford} for more details). A \textit{general model} is a pair $ \M=\langle M,G\rangle$, where $M=\langle D,I\rangle$ is a model and $G$ is a set of relations on $D$, each of which is a subset of $D^n$ for some $n \geq 1$; that is, $ G \subseteq \bigcup_{n \geq 1}^{}\mathcal{P}(D^n) $. 
 Notice that $v(X) \in G \subseteq \mathcal{P}(D^n)$. We define the notion of satisfaction of a formula $\varphi$ with $v$ in a general model, symbolically $\M,v\models \varphi$, for second-order formulas as follows:
\[
\begin{aligned}
\M, v \models X(t_1, \ldots, t_n) &\text{ iff } \langle v(t_1), \ldots, v(t_n) \rangle \in v(X), \quad \text{if $X$ is $n$-ary}, \\
\M, v \models X = Y &\text{ iff } v(X) = v(Y), \\
\M, v \models (\lambda X \psi)\iota Y\varphi & \text{ iff } \text{there is an } O \in G \text{ such that } \M, v_O^X \models \psi, \\
& \quad \M, v_O^X \models \varphi^Y_X, \text{ and for any } Y\text{-variant } v' \text{ of } v_O^X, \\
& \quad \text{if } \M, v' \models \varphi, \text{ then } v'(Y) = O \\
\M, v \models \forall X \varphi & \text{ iff } \M, v_O^X \models \varphi, \text{ for all } O \in G,\\
\M, v \models \exists X \varphi & \text{ iff } \M, v_O^X \models \varphi, \text{ for some } O \in G.
\end{aligned}
\]

Notice that each relational constant \( K^n \) corresponds to a set of \( n \)-tuples of individual constants, and is interpreted as an element of the general model domain \( G \subseteq \mathcal{P}(D^n) \). By this concept, we understand the elements of \( G \) to be syntactically named via relational constants, allowing us to treat them as concrete surrogates for second-order values during the construction of canonical models.

 The notions of a satisfiable formula and a valid formula are defined in a standard way.  The consequence relation is defined as follows, for all $\Gamma\subseteq\Fii$ and $\varphi\in\Fii$: $\Gamma\models_\RLL \varphi$ iff in every general model $\M$ and every assignment $v$, if $\M,v\models \psi$, for all $\psi\in\Gamma$, then $\M,v\models \varphi$. This can be extended to the multiple-conclusion consequence relation: for all $\Gamma,\Delta\subseteq\Fii$, $\Gamma\models_\RLL \Delta$ iff in every general model $\M$ and every assignment $v$, if $\M,v\models \psi$, for all $\psi\in\Gamma$, then $\M,v\models \chi$, for some $\chi\in\Delta$. 
A sequent is an ordered pair written as $\Gamma\Rightarrow\Delta$, where $\Gamma$ and $\Delta$ are finite multisets of formulas. We write $\models_\RLL \Gamma\Rightarrow\Delta$ iff $\Gamma\models_\RLL \Delta$. If $\mathcal{S}$ is a set of sequents and $H$ is a sequent, we write $\mathcal{S} \models_{\RLL} H$ iff $\models_\RLL S$, for all $S\in\mathcal{S}$, implies $\models_\RLL H$.

\begin{figure}[t]
    \centering
\begin{align*}
&\textrm{(Cut)} \; \frac{\Gamma \Rightarrow \Delta, \varphi \quad \varphi, \Pi \Rightarrow \Sigma}{\Gamma, \Pi \Rightarrow \Delta, \Sigma} 
\quad 
\textrm{(AX)} \; \varphi \Rightarrow \varphi 
\quad
\textrm{(W$\Rightarrow$)} \; \frac{\Gamma \Rightarrow \Delta}{\varphi, \Gamma \Rightarrow \Delta}
\quad
\\[1em]
& \textrm{($\Rightarrow$W)} \; \frac{\Gamma \Rightarrow \Delta}{ \Gamma \Rightarrow \Delta,\varphi}
\quad 
\textrm{(C$\Rightarrow$)} \; \frac{\varphi, \varphi, \Gamma \Rightarrow \Delta}{\varphi, \Gamma \Rightarrow \Delta}
\quad\textrm{($\Rightarrow$C)} \; \frac{\Gamma \Rightarrow \Delta,\varphi, \varphi}{\Gamma \Rightarrow \Delta,\varphi}
\\[1em]
& \textrm{($\land\Rightarrow$)} \; \frac{\varphi, \psi, \Gamma \Rightarrow \Delta}{\varphi \land \psi, \Gamma \Rightarrow \Delta}
\quad
\textrm{($\Rightarrow\land$)} \; \frac{\Gamma \Rightarrow \Delta, \varphi \quad \Gamma \Rightarrow \Delta, \psi}{\Gamma \Rightarrow \Delta, \varphi \land \psi}
\quad
\textrm{($\neg\Rightarrow$)} \; \frac{\Gamma \Rightarrow \Delta, \varphi}{\neg \varphi,\Gamma \Rightarrow \Delta}
\\[1em]
&\textrm{($\lor\Rightarrow$)} \; \frac{\varphi, \Gamma \Rightarrow \Delta \quad \psi, \Gamma \Rightarrow \Delta}{\varphi \lor \psi, \Gamma \Rightarrow \Delta}
\quad 
\textrm{($\Rightarrow\lor$)} \; \frac{\Gamma \Rightarrow \Delta, \varphi, \psi}{\Gamma \Rightarrow \Delta, \varphi \lor \psi}
\quad
\textrm{($\Rightarrow\neg$)} \; \frac{\varphi, \Gamma \Rightarrow \Delta}{\Gamma \Rightarrow \Delta, \neg \varphi}
\\[1em]
&\textrm{($\to\Rightarrow$)} \; \frac{\Gamma \Rightarrow \Delta, \varphi \quad \psi, \Gamma \Rightarrow \Delta}{\varphi \to \psi,\Gamma \Rightarrow \Delta}
\quad 
\textrm{($\Rightarrow\to$)} \; \frac{\varphi,\Gamma \Rightarrow \Delta,  \psi}{\Gamma \Rightarrow \Delta,\varphi \to \psi}
\quad (\forall \Rightarrow) \; \frac{\varphi^x_b, \Gamma \Rightarrow \Delta}{\forall x \varphi, \Gamma \Rightarrow \Delta} 
\\[1em]
&(\leftrightarrow\Rightarrow)  \; \frac{\Gamma \Rightarrow \Delta, \varphi, \psi \quad \varphi, \psi, \Gamma \Rightarrow \Delta}{\varphi \leftrightarrow \psi, \Gamma \Rightarrow \Delta} 
\quad
(\Rightarrow \forall) \; \frac{\Gamma \Rightarrow \Delta, \varphi^x_a}{\Gamma \Rightarrow \Delta, \forall x \varphi}
\quad 
\textrm{($\Rightarrow\exists$)} \; \frac{\Gamma \Rightarrow \Delta, \varphi^x_b}{\Gamma \Rightarrow \Delta, \exists x \varphi}
\\[1em]
&(\Rightarrow \leftrightarrow) \; \frac{\varphi, \Gamma \Rightarrow \Delta, \psi \quad \psi, \Gamma \Rightarrow \Delta, \varphi}{\Gamma \Rightarrow \Delta, \varphi \leftrightarrow \psi} 
\quad
\textrm{($\exists\Rightarrow$)} \; \frac{\varphi^x_a, \Gamma \Rightarrow \Delta}{\exists x \varphi, \Gamma \Rightarrow \Delta}
\quad
\textrm{($= +$)} \; \frac{b = b, \Gamma \Rightarrow \Delta}{\Gamma \Rightarrow \Delta} 
\\[1em]
&\textrm{($= -$)} \; \frac{\mathscr{A}^x_c, \Gamma \Rightarrow \Delta}{b = c, \mathscr{A}^x_b, \Gamma \Rightarrow \Delta} 
\quad 
\textrm{($\lambda \Rightarrow$)} \; \frac{\psi^x_b, \Gamma \Rightarrow \Delta}{(\lambda x \psi) b, \Gamma \Rightarrow \Delta}
\quad \textrm{($\Rightarrow \lambda$)} \; \frac{\Gamma \Rightarrow \Delta, \psi^x_b}{\Gamma \Rightarrow \Delta, (\lambda x \psi) b}
\\[1em]
&\textrm{($\iota_1 \Rightarrow$)} \; \frac{\varphi^y_a, \psi^x_a, \Gamma \Rightarrow \Delta}{(\lambda x \psi)\iota y \varphi, \Gamma \Rightarrow \Delta}
\quad
\textrm{($\iota_2 \Rightarrow$)} \; \frac{\Gamma \Rightarrow \Delta, \varphi^y_b \quad \Gamma \Rightarrow \Delta, \varphi^y_c\quad b = c, \Gamma \Rightarrow \Delta}{(\lambda x \psi)\iota y \varphi, \Gamma \Rightarrow \Delta} 
\\[1em]
&\textrm{($\Rightarrow \iota$)} \; \frac{\Gamma \Rightarrow \Delta, \varphi^y_b \quad \Gamma \Rightarrow \Delta, \psi^x_b \quad \varphi^y_a, \Gamma \Rightarrow \Delta, a = b}{\Gamma \Rightarrow \Delta, (\lambda x \psi)\iota y \varphi}
\end{align*}
where \(a\) is a fresh parameter (\textit{Eigenvariable}), not present in \(\Gamma, \Delta\) and \(\varphi\), whereas \(b, c\) are arbitrary parameters. $\mathscr{A}$ in \((= -)\) is an atomic formula.
    \caption{Sequent calculus for $\RL$.}
    \label{RL}
\end{figure}

\section{Sequent calculi. Soundness, Completeness, and Cut admissibility}\label{ND}
On Figures \ref{RL} and \ref{RLL}, we present sequent calculi for $\RL$ and $\RLL$, respectively. The sequent calculus for $\RL$ is due to Indrzejczak and K\"{u}rbis \cite{IKRL}; it is an extension of the calculus G1c of Troelstra
and Schwichtenberg \cite{TroelstraSchwichtenberg} by rules for identity and lambda atoms.

If $\mathcal{S}$ is a set of sequents and $H$ is a sequent, we write $\mathcal{S} \vdash_{\RLL} H$ iff there is a proof of $H$ from $ \mathcal{S} $ in $ \RLL $, i.e. there exists a tree whose nodes are sequents such that the leaves are either axioms or members of $\mathcal{S}$, the root is $H$, and each node is obtained from its immediate predecessors by an application of a rule from the calculus. 
If $\mathcal{S}$ is a set of sequents and $H$ is a sequent, then we write $\mathcal{S} \vdash^{cf}_{\RLL} H$ iff $\mathcal{S} \vdash_{\RLL} H$ and each cut is on a formula that belongs to $\mathcal{S}$. Two examples of proofs in $ \RLL $ are presented in the proof of the subsequent proposition.

\begin{proposition}\label{One}
    The following rules are derivable (with the help of the cut rule) in the sequent calculus for $\RLL$, where $\mathscr{A}$ is an atomic formula\textup{:}
\begin{center}
$(=^2 +)$ \; $\dfrac{X = X, \Gamma \Rightarrow \Delta}{\Gamma \Rightarrow \Delta} $
\quad 
$(=^2 -)$ \; $\dfrac{\mathscr{A}^X_C, \Gamma \Rightarrow \Delta}{B = C, \mathscr{A}^X_B, \Gamma \Rightarrow \Delta} $ 
\end{center}    
\end{proposition}
\begin{proof}
Consider the following proofs in $ \RLL $.
{\small 
\begin{prooftree}
\EnableBpAbbreviations
\AXC{$ X^{x_1,\ldots,x_n}_{a_1,\ldots,a_n}\Rightarrow X^{x_1,\ldots,x_n}_{a_1,\ldots,a_n} $}
\AXC{$ X^{x_1,\ldots,x_n}_{a_1,\ldots,a_n}\Rightarrow X^{x_1,\ldots,x_n}_{a_1,\ldots,a_n} $}
\RL{$ (\Rightarrow =^2) $}
\BIC{$ \Rightarrow X=X $}
\AXC{$ X = X, \Gamma \Rightarrow \Delta $}
\LL{(Cut)}
\BIC{$ \Gamma \Rightarrow \Delta $}
\end{prooftree}
}

 \begin{prooftree}
\EnableBpAbbreviations
\AXC{$ \mathscr{A}^X_B \Rightarrow \mathscr{A}^X_B $}
\RL{($ \Rightarrow $W)}
\UIC{$ \mathscr{A}^X_B \Rightarrow \mathscr{A}^X_B,\mathscr{A}^X_C $}
\AXC{$ \mathscr{A}^X_C \Rightarrow \mathscr{A}^X_C $}
\RL{(W$ \Rightarrow $)}
\UIC{$ \mathscr{A}^X_C,\mathscr{A}^X_B \Rightarrow \mathscr{A}^X_C $}
\RL{$ (=^2\Rightarrow) $}
\BIC{$ B = C, \mathscr{A}^X_B \Rightarrow \mathscr{A}^X_C $}
\AXC{$ \mathscr{A}^X_C, \Gamma \Rightarrow \Delta $}
\LL{(Cut)}
\BIC{$ B = C, \mathscr{A}^X_B, \Gamma \Rightarrow \Delta $}
\end{prooftree}
\end{proof}

\begin{proposition}\cite[Lemma 1]{IKRL}
The following two statements hold in $ \RL $:
\begin{itemize}\itemsep=0pt
\item $\vdash_{\RL} b_1=b_2,\varphi^x_{b_1}\Rightarrow\varphi^x_{b_2} $, for any formula $ \varphi $, 
\item if $ \vdash_{\RL}^n\Gamma\Rightarrow\Delta $, then $ \vdash_{\RL}^n\Gamma^{b_1}_{b_2}\Rightarrow\Delta^{b_1}_{b_2} $, where $ n $ is the height of a proof.
\end{itemize}
\end{proposition}

\begin{proposition}
The following two statements hold in $ \RLL $:
\begin{itemize}\itemsep=0pt
\item $\vdash_{\RLL} B_1=B_2,\varphi^X_{B_1}\Rightarrow\varphi^X_{B_2} $, for any formula $ \varphi $, 
\item if $ \vdash_{\RLL}^n\Gamma\Rightarrow\Delta $, then $ \vdash_{\RLL}^n\Gamma^{B_1}_{B_2}\Rightarrow\Delta^{B_1}_{B_2} $, where $ n $ is the height of a proof.
\end{itemize}
\end{proposition}
\begin{proof}
Similarly to \cite[Lemma 1]{IKRL}. 
\end{proof}

\begin{figure}[t]
    \centering
\begin{align*}
&(=^2\Rightarrow)  \; \frac{\Gamma \Rightarrow \Delta, X^{x_1,\ldots,x_n}_{b_1,\ldots,b_n}, Y^{x_1,\ldots,x_n}_{b_1,\ldots,b_n} \quad X^{x_1,\ldots,x_n}_{b_1,\ldots,b_n}, Y^{x_1,\ldots,x_n}_{b_1,\ldots,b_n}, \Gamma \Rightarrow \Delta}{X = Y, \Gamma \Rightarrow \Delta} 
\\[1em]
&(\Rightarrow =^2) \; \frac{X^{x_1,\ldots,x_n}_{a_1,\ldots,a_n}, \Gamma \Rightarrow \Delta, Y^{x_1,\ldots,x_n}_{a_1,\ldots,a_n} \quad Y^{x_1,\ldots,x_n}_{a_1,\ldots,a_n}, \Gamma \Rightarrow \Delta, X^{x_1,\ldots,x_n}_{a_1,\ldots,a_n}}{\Gamma \Rightarrow \Delta,X=Y}\\[1em]
&\textrm{($\exists^2\Rightarrow$)} \; \frac{\varphi^X_A, \Gamma \Rightarrow \Delta}{\exists X \varphi, \Gamma \Rightarrow \Delta}
\quad
\textrm{($\Rightarrow\exists^2$)} \; \frac{\Gamma \Rightarrow \Delta, \varphi^X_B}{\Gamma \Rightarrow \Delta, \exists X \varphi}
\quad
\textrm{($\iota_1^2 \Rightarrow$)} \; \frac{\varphi^Y_A, \psi^X_A, \Gamma \Rightarrow \Delta}{(\lambda X \psi)\iota Y \varphi, \Gamma \Rightarrow \Delta}
\\[1em]
&(\forall^2 \Rightarrow) \; \frac{\varphi^X_B, \Gamma \Rightarrow \Delta}{\forall X \varphi, \Gamma \Rightarrow \Delta}
\quad
\textrm{($\iota_2^2 \Rightarrow$)} \; \frac{\Gamma \Rightarrow \Delta, \varphi^Y_B \quad \Gamma \Rightarrow \Delta, \varphi^Y_C\quad B = C, \Gamma \Rightarrow \Delta}{(\lambda X \psi)\iota Y \varphi, \Gamma \Rightarrow \Delta} 
\\[1em]
&(\Rightarrow \forall^2) \; \frac{\Gamma \Rightarrow \Delta, \varphi^X_A}{\Gamma \Rightarrow \Delta, \forall X \varphi}
\quad
\textrm{($\Rightarrow \iota^2$)} \; \frac{\Gamma \Rightarrow \Delta, \varphi^Y_B \quad \Gamma \Rightarrow \Delta, \psi^X_B \quad \varphi^Y_A, \Gamma \Rightarrow \Delta, A = B}{\Gamma \Rightarrow \Delta, (\lambda X \psi)\iota Y \varphi}
\\[1em]
\end{align*}
where \(a_1,\ldots,a_n\) are fresh individual parameters, not present in \(\Gamma\) and \(\Delta\); $b_1,\ldots,b_n$ are arbitrary individual parameters; \(A\) is a fresh relational parameter, not present in \(\Gamma, \Delta\) and \(\varphi\); \(B\) and \(C\) are arbitrary relational parameters.
    \caption{$\RLL$ extends $\RL$ by the following second-order rules.}
    \label{RLL}
\end{figure}

\begin{theorem}
All the rules of $\RLL$ are sound. 
\end{theorem}
\begin{proof}
    By a routine check. For the case of $\RL$ see \cite{IKRL}.
\end{proof}

As for our completeness proof, we follow the Hintikka-style strategy originally described in \cite{AvronLahav,LahavAvron} for hypersequent calculi for some non-classical first-order logics. We adapt this proof for the second-order case. Readers interested in further details on the Hintikka-style approach to completeness proofs may consult \cite{IndrzejST,IndrzejTense}. Notice that the Henkin-style completness proof for $\RL$ is given in \cite{IKRL}. 

\begin{definition}\cite[Definition 4.9]{AvronLahav}
An \emph{extended sequent} is an ordered pair of (possibly infinite) sets of formulas. Given two extended sequents $S_1 = \Gamma_1\Rightarrow \Delta_1$ and $S_2 =  \Gamma_2\Rightarrow \Delta_2$, we write $S_1 \sqsubseteq S_2$ if $\Gamma_1 \subseteq \Gamma_2$ and $\Delta_1\subseteq \Delta_2$. An extended sequent is called \emph{finite} if it consists of finite sets of formulas.
\end{definition}

\begin{definition}\cite[Definition 4.11; extended for the second-order quantifiers and $=$, $\iota$, and $\lambda$]{AvronLahav}
An extended sequent $\Gamma \Rightarrow \Delta$ admits \emph{the witness property} if the following hold (recall that parameters play the role of free variables):
\begin{enumerate}[$(1)$]
    \item If $\forall x\varphi \in \Delta$, then $\varphi^x_k \in \Delta$, for some individual constant $k$.
    \item If $\exists x\varphi \in \Gamma$, then $\varphi^x_k \in \Gamma$, for some individual constant $k$.
     \item If $\forall X\varphi \in \Delta$, then $\varphi^X_K \in \Delta$, for some relational constant $K$.
    \item If $\exists X\varphi \in \Gamma$, then $\varphi^X_K \in \Gamma$, for some relational constant $K$. 
\item If $(\lambda x\psi) \iota y \varphi \in \Gamma$, then $\varphi^y_k \in \Gamma$ and $\psi^x_k \in \Gamma$, for some individual constant $k$.
\item If $(\lambda x\psi) \iota y \varphi \in \Delta$, then for each individual parameter $b$, $\varphi^y_b\in \Delta$, or
$\psi^x_b, \in \Delta$, or for some individual constant $k$, $k=b \in \Delta$
and $\varphi^y_k \in \Gamma$.
\item If $(\lambda X\psi) \iota Y \varphi \in \Gamma$, then $\varphi^Y_K,\psi^X_K \in \Gamma$, for some relational constant $K$.
\item If $(\lambda X\psi) \iota Y \varphi \in \Delta$, then for  each relational parameter $B$, $\varphi^Y_B \in \Delta$, or $\psi^X_B \in \Delta$, or for some relational constant $K$, $K=B \in \Delta$ and $\varphi^Y_K \in \Gamma$.
\item If $X=Y\in\Delta$, then either \big($X^n(k_1, \ldots, k_n)\in \Gamma$ and $Y^n(k_1, \ldots, k_n)\in\Delta$\big) or \big($Y^n(k_1, \ldots, k_n)\in \Gamma$ and $X^n(k_1, \ldots, k_n)\in \Delta$\big), for some individual constants $k_1,\ldots,k_n$.
\end{enumerate}
\end{definition}

\begin{definition}\cite[Definition 4.12; adapted for the case of ordinary sequents]{AvronLahav}
 Let $\Gamma \Rightarrow \Delta$ be an extended sequent and $\mathcal{S}$ be a set of sequents. 
\begin{enumerate}[$(1)$]
    \item $\Gamma \Rightarrow \Delta$ is called \textit{$\mathcal{S}$-consistent} if $\mathcal{S} \not\vdash^{cf}_{\RLL} H$, for every sequent $H\sqsubseteq\Gamma \Rightarrow \Delta$.
    \item $\Gamma \Rightarrow \Delta$ is \textit{internally $\mathcal{S}$-maximal with respect to an $\mathscr{L}$-formula $\varphi$} iff:
    \begin{enumerate}[$(a)$]
        \item If $\varphi \not\in \Gamma$, then $\Gamma, \varphi \Rightarrow \Delta$ is not $\mathcal{S}$-consistent.
        \item If $\varphi \not\in \Delta$, then $\Gamma \Rightarrow \Delta, \varphi$ is not $\mathcal{S}$-consistent.
    \end{enumerate}
    \item $\Gamma \Rightarrow \Delta$ is called \textit{internally $\mathcal{S}$-maximal} if it is internally $\mathcal{S}$-maximal with respect to any  $\mathscr{L}$-formula.
    \item $\Gamma \Rightarrow \Delta$ is called \textit{$\mathcal{S}$-maximal} if it is $\mathcal{S}$-consistent, internally $\mathcal{S}$-maximal, and it admits the witness property.
\end{enumerate}
\end{definition}

\begin{lemma}\cite[Lemma 4.13]{AvronLahav}\cite[Proposition 41]{LahavAvron}\label{Prop41}
\textit{Let $\Gamma \Rightarrow \Delta$ be an extended sequent that is internally $\mathcal{S}$-maximal with respect to an $\mathscr{L}$-formula $\varphi$. Then:}
\begin{enumerate}[$(1)$]
    \item\textit{If $\varphi \not\in \Gamma$, then $\mathcal{S} \vdash^{cf}_{\RLL}  \Theta, \varphi \Rightarrow \Lambda$ for some sequent $\Theta \Rightarrow \Lambda\sqsubseteq \Gamma \Rightarrow \Delta$.}
    \item \textit{If $\varphi \not\in \Delta$, then $\mathcal{S} \vdash^{cf}_{\RLL} \Theta \Rightarrow \Lambda, \varphi$ for some sequent $\Theta \Rightarrow \Lambda\sqsubseteq \Gamma \Rightarrow \Delta$.}
    \end{enumerate}
\end{lemma}
\begin{proof}
    By an adaptation of the proof of Proposition 41 from \cite{LahavAvron}.
\end{proof}

\begin{lemma}
\cite[Lemma 4.15]{AvronLahav}\cite[Lemma 43]{LahavAvron} \textit{Let $\mathcal{S}$ be a set of sequents and let}
 $\Gamma \Rightarrow \Delta $
\textit{be a $\mathcal{S}$-consistent finite extended sequent. Then, there exists a $\mathcal{S}$-consistent finite extended sequent}
$\Gamma' \Rightarrow \Delta'$ 
\textit{such that $\Gamma \subseteq \Gamma'$ and $\Delta \subseteq \Delta'$  and $\Gamma' \Rightarrow \Delta'$, and $\Gamma' \Rightarrow \Delta'$ admits the witness property.}
\end{lemma}
\begin{proof}
Similarly to \cite[Lemma 43]{LahavAvron}. If $\forall x\varphi\in\Delta$, we take a fresh individual constant $k$ and add the formula $\varphi^x_k$ to $\Delta$. If $\exists x\varphi\in\Gamma$, we again take a fresh constant $k$ and add the
formula $\varphi^x_k$ to $\Gamma$. Similarly, we proceed for the second-order quantification, $=$, $\iota$, and $\lambda$. We continue this procedure until the obtained extended sequent admits the witness property. 
Since the set of formulas in the sequent 
\( \Gamma \Rightarrow \Delta \) is finite, and at each step the added formula is either a subformula or a witness-instantiated variant of an existing formula~-- where witness instantiations are limited to a finite set of constants and predicate names~-- the number of possible new formulas is finite as well. Furthermore, formulas are added only once. Therefore, the saturation procedure must terminate after a finite number of steps. 
The finite extended sequent \( \Gamma' \Rightarrow \Delta' \) is produced from \( \Gamma \Rightarrow \Delta \) through this method. We will demonstrate that each such extension preserves the \( \mathcal{S} \)-consistency of the extended sequent.

Suppose that $(\lambda X\psi) \iota Y \varphi \in \Delta$. Let $\Gamma' \Rightarrow \Delta'$ be the extended sequent obtained from $\Gamma\Rightarrow \Delta$ 
by adding $\varphi^Y_B$,  $\psi^X_B$, and  $K=B$ to $\Delta$ and $\varphi^Y_K$ to $\Gamma$, where $K$ is a relational constant that does not occur in $\Gamma\Rightarrow \Delta$.  
Assume for the contradiction that $\Gamma' \Rightarrow \Delta'$ is not $\mathcal{S}$-consistent. By Lemma \ref{Prop41}, there are sequents $\Theta \Rightarrow \Lambda,\Pi\Rightarrow\Sigma,\Upsilon\Rightarrow\Phi,\Psi\Rightarrow\Omega\sqsubseteq \Gamma \Rightarrow \Delta$ such that 
$\mathcal{S}\vdash^{cf}_{\RLL}\Theta\Rightarrow\Lambda,\varphi^Y_B$, 
$\mathcal{S}\vdash^{cf}_{\RLL}\Pi\Rightarrow\Sigma,\psi^X_B$, 
$\mathcal{S}\vdash^{cf}_{\RLL}\varphi^Y_K,\Upsilon\Rightarrow\Phi$, and $\mathcal{S}\vdash^{cf}_{\RLL}\Psi\Rightarrow\Omega,K=B$. By weakening, we get $\mathcal{S}\vdash^{cf}_{\RLL}\varphi^Y_K,\Upsilon\Rightarrow\Phi,K=B$ (or one could get $\mathcal{S}\vdash^{cf}_{\RLL}\varphi^Y_K,\Psi\Rightarrow\Omega,K=B$). 
 Applying weakening and $(\Rightarrow \iota^2)$, we obtain
$\mathcal{S}\vdash^{cf}_{\RLL}\Theta,\Pi,\Upsilon\Rightarrow\Lambda,\Sigma,\Phi,(\lambda X\psi) \iota Y \varphi$. 
This contradicts the fact that $\Gamma\Rightarrow \Delta$ is $\mathcal{S}$-consistent.

The other cases are considered similarly.
\end{proof}

\begin{lemma}
\cite[Lemma 4.16]{AvronLahav}\cite[Lemma 44]{LahavAvron}
\textit{Let $\mathcal{S}$ be a set of sequents and}
$H$ 
\textit{be a $\mathcal{S}$-consistent finite extended sequent. Let $\varphi$ be an $\mathcal{L}$-formula. Then, there exists a $\mathcal{S}$-consistent finite extended sequent $H'$ such that:}
\begin{itemize}
    \item $H \subseteq H' $,
    \item $H'$ is internally $\mathcal{S}$-maximal with respect to $\varphi$,
    \item $H'$ admits the witness property.
\end{itemize}
\end{lemma}
\begin{proof}
    Similarly to the proof of Lemma 44 from \cite{LahavAvron}.
\end{proof}

\begin{lemma}\cite[Lemma 4.17]{AvronLahav}\cite[Lemma 45]{LahavAvron}\label{Lemma47}
Let $\mathcal{S}$ be a set of sequents. Every $\mathcal{S}$-consistent sequent can be extended to a $\mathcal{S}$-maximal extended sequent $\Gamma \Rightarrow \Delta$.
\end{lemma}
\begin{proof}
    Similarly to the proof of Lemma 45 from \cite{LahavAvron}.
\end{proof}

\begin{theorem}[strong completeness of $\RLL$]\label{Compl}
    Let $\mathcal{S}$ be a set of sequents and $H$ be a sequent. If $\mathcal{S} \models_{\RLL} H$, then $\mathcal{S} \vdash^{cf}_{\RLL} H$.
\end{theorem}
\begin{proof}
Suppose that $\mathcal{S} \not\vdash^{cf}_{\RLL} H$. By Lemma \ref{Lemma47}, there exists a $\mathcal{S}$-maximal extended sequent $\Gamma \Rightarrow\Delta$ such that $H\sqsubseteq \Gamma \Rightarrow \Delta$. Using $\Gamma \Rightarrow \Delta$, we construct a structure $M=\langle D, I \rangle$. $D$ is the set of equivalence classes of terms. We denote the equivalence class to which a term $t$ belongs by $[t]$. 
For all individual variables $v(x) = [x]$, for all individual parameters $v(a) = [a]$, for all individual constants, $I(k) = [k]$. For all predicate letters (including $=$), $\langle [t_1], \ldots, [t_n] \rangle \in I(P^n) \text{ iff } P^n(t_1, \ldots, t_n) \in \Gamma$. 
We construct a general structure $\M=\langle M, G \rangle$ as follows: $M$ is defined above and $G$ is a set of subsets on $D$. As for relational variables, parameters, and constants, we postulate that $v(X)\in G$, $v(A)\in G$, and $I(K)\in G$. For all relational variables, $\langle [t_1], \ldots, [t_n] \rangle \in v(X^n) \text{ iff } X^n(t_1, \ldots, t_n) \in \Gamma$. 
We prove the following two statements together by induction on the complexity of $\chi$:
\begin{enumerate}[(a)]
    \item If $\chi \in \Gamma$ then $\M, v \models \chi$.
    \item If $\chi \in \Delta$ then $\M, v \not\models \chi$.
\end{enumerate}

\underline{$\chi$ is $t_1 = t_2$.} (a) $\M, v \models t_1 = t_2$ 
    iff $v(t_1) = v(t_2)$
    iff $[t_1] = [t_2]$ 
    and as these are equivalence classes, 
    iff $t_1 = t_2 \in \Gamma$. (b) If $t_1 = t_2 \in \Delta$, $t_1 = t_2 \not\in \Gamma$, since $\mathcal{S} \not\vdash^{cf}_{\RLL} H$. Then $[t_1]\not = [t_2]$. Hence, $v(t_1)\not = v(t_2)$. Thus, $\M, v \models t_1 \not= t_2$.

\underline{$\chi$ is $X^n(t_1, \ldots, t_n)$.} (a) $\M, v \models X^n(t_1, \ldots, t_n)$ 
    iff $\langle [t_1], \ldots, [t_n] \rangle \in v(X^n)$ 
    iff $X^n(t_1, \ldots, t_n) \in \Gamma$. (b) If $X^n(t_1, \ldots, t_n) \in \Delta$, then $X^n(t_1, \ldots, t_n) \not\in \Gamma$, since $\mathcal{S} \not\vdash^{cf}_{\RLL} H$. Then $\langle [t_1], \ldots, [t_n] \rangle \not\in v(X^n)$. Hence, $\M, v \not\models X^n(t_1, \ldots, t_n)$.

\underline{$\chi$ is $\varphi\leftrightarrow \psi$.} (a) Assume that $\varphi\leftrightarrow\psi\in \Gamma $. Suppose that ($\varphi\not\in \Gamma$ or $\psi\not\in \Gamma$) and ($\varphi\not\in \Delta$ or $\psi\not\in \Delta$). By Lemma \ref{Prop41}, there are sequents $\Theta \Rightarrow \Lambda$ and $\Pi\Rightarrow\Sigma$ such that $\Theta \Rightarrow \Lambda,\Pi\Rightarrow\Sigma\sqsubseteq\Gamma \Rightarrow\Delta$ as well as $\mathcal{S} \vdash^{cf}_{\RLL}\varphi,\psi, \Theta \Rightarrow \Lambda$ and $\mathcal{S} \vdash^{cf}_{\RLL} \Pi\Rightarrow\Sigma,\varphi,\psi$.
 By weakening and ($\leftrightarrow\Rightarrow$), $\mathcal{S} \vdash^{cf}_{\RLL}\varphi\leftrightarrow\psi, \Theta,\Pi \Rightarrow \Lambda,\Sigma$.
This contradicts the $\mathcal{S}$-consistency of $\Gamma \Rightarrow\Delta$. 
Hence, either $\varphi,\psi\in \Gamma$ or $\varphi,\psi\in \Delta$. Thus, by the induction hypothesis, either ($\M, v \models \varphi$ and $\M, v \models \psi$) or ($\M, v \not\models \varphi$ and $\M, v \not\models \psi$). Therefore, $\M, v \models \varphi\leftrightarrow \psi$.

(b) Assume that $\varphi\leftrightarrow\psi\in \Delta $. Suppose that ($\varphi\not\in \Gamma$ or $\psi\not\in\Delta$) and ($\psi\not\in \Gamma$ or $\varphi\not\in \Delta$). By Lemma \ref{Prop41}, there are sequents $\Theta \Rightarrow \Lambda$ and $\Pi\Rightarrow\Sigma$ such that $\Theta \Rightarrow \Lambda,\Pi\Rightarrow\Sigma\sqsubseteq\Gamma \Rightarrow\Delta$ as well as 
$\mathcal{S} \vdash^{cf}_{\RLL}\varphi, \Theta \Rightarrow \Lambda,\psi$ and $\mathcal{S} \vdash^{cf}_{\RLL} \psi,\Pi\Rightarrow\Sigma,\varphi$.
 By weakening and ($\Rightarrow\leftrightarrow$), $\mathcal{S} \vdash^{cf}_{\RLL} \Theta,\Pi \Rightarrow \Lambda,\Sigma,\varphi\leftrightarrow\psi$.
This contradicts the $\mathcal{S}$-consistency of $\Gamma \Rightarrow\Delta$. 
Hence, either ($\varphi\in \Gamma$ and $\psi\in\Delta$) or ($\psi\in \Gamma$ and $\varphi\in \Delta$). Thus, by the induction hypothesis, either ($\M, v \models \varphi$ and $\M, v \not\models \psi$) or ($\M, v \not\models \varphi$ and $\M, v \models \psi$). Therefore, $\M, v \not\models \varphi\leftrightarrow \psi$.

\underline{$\chi $ is $ \exists X \varphi$.} 
 (a)   Assume that $\exists X \varphi \in \Gamma$. By the witness property of $\Gamma \Rightarrow\Delta$, there exists a relational constant $K$ such that $\varphi^X_K \in \Gamma$. By the induction hypothesis, $\M,v \models \varphi^X_K$. Since $K\in G$, it holds that $\M, v \models \exists X \varphi$.
        
(b) Assume that $\M,v \models \exists X \varphi$. We show that $\exists X \varphi \notin \Delta$. By definition, there exists some $O \in G$ such that $\M,v \models \varphi^X_O$. By the induction hypothesis, $\varphi^X_O \notin \Delta$. 
By Lemma \ref{Prop41}, there exist a sequent $\Theta \Rightarrow\Lambda \sqsubseteq \Gamma \Rightarrow \Delta$ such that $\mathcal{S} \vdash^{cf}_{\RLL} \Theta \Rightarrow\Lambda,\varphi^X_O$. By the rule ($\Rightarrow\exists^2$), $\mathcal{S} \vdash^{cf}_{\RLL} \Theta \Rightarrow\Lambda,\exists X \varphi$. 
 Since $\Gamma \Rightarrow \Delta$ is $\mathcal{S}$-consistent, $\exists X \varphi \notin \Delta$.

\underline{$\chi$ is $X=Y$.} (a) Assume that $X=Y\in \Gamma $. Suppose that there are $o_1, \ldots, o_n\in D$ such that ($X^n(o_1, \ldots, o_n)\not\in \Gamma$ or $Y^n(o_1, \ldots, o_n)\not\in \Gamma$) and ($X^n(o_1, \ldots, o_n)\not\in \Delta$ or $Y^n(o_1, \ldots, o_n)\not\in \Delta$). By Lemma \ref{Prop41}, there are sequents 
$\Theta \Rightarrow \Lambda,\Pi\Rightarrow\Sigma\sqsubseteq\Gamma \Rightarrow\Delta$ such that $\mathcal{S} \vdash^{cf}_{\RLL}X^n(o_1, \ldots, o_n),Y^n(o_1, \ldots, o_n), \Theta \Rightarrow \Lambda$ and $\mathcal{S} \vdash^{cf}_{\RLL} \Pi\Rightarrow\Sigma,X^n(o_1, \ldots, o_n),Y^n(o_1, \ldots, o_n)$.
 By the rules of weakening and ($=^2\Rightarrow$), $\mathcal{S} \vdash^{cf}_{\RLL}X=Y, \Theta,\Pi \Rightarrow \Lambda,\Sigma$.
This contradicts the $\mathcal{S}$-consistency of $\Gamma \Rightarrow\Delta$. Therefore, for all $o_1, \ldots, o_n\in D$, either \big($X^n(o_1, \ldots, o_n)\in \Gamma$ and $Y^n(o_1, \ldots, o_n)\in \Gamma$\big) or \big($X^n(o_1, \ldots, o_n)\in \Delta$ and $Y^n(o_1, \ldots, o_n)\in \Delta$\big). Thus, by the induction hypothesis, either \big($\M,v \models X^n(o_1, \ldots, o_n)$ and $\M,v \models Y^n(o_1, \ldots, o_n)$\big) or \big($\M,v \not\models X^n(o_1, \ldots, o_n)$ and $\M,v \not\models Y^n(o_1, \ldots, o_n)$\big). Therefore, it holds that either \big($\langle [o_1], \ldots, [o_n] \rangle \in v(X^n)$ and $\langle [o_1], \ldots, [o_n] \rangle \in v(Y^n)$\big) or \big($\langle [o_1], \ldots, [o_n] \rangle\not \in v(X^n)$ and $\langle [o_1], \ldots, [o_n] \rangle \not\in v(Y^n)$\big). Hence, $v(X) = v(Y)$. Thus, 
$\M,v \models X=Y$. 

(b) Assume that $X=Y\in \Delta $. By the witness property of $\Gamma \Rightarrow\Delta$, 
there are individual constants $k_1,\ldots,k_n\in D$ such that
either \big($X^n(k_1, \ldots, k_n)\in \Gamma$ and $Y^n(k_1, \ldots, k_n)\in\Delta$\big) or \big($Y^n(k_1, \ldots, k_n)\in \Gamma$ and $X^n(k_1, \ldots, k_n)\in \Delta$\big). Thus, by the induction hypothesis, either \big($\M, v \models X^n(k_1, \ldots, k_n)$ and $\M, v \not\models Y^n(k_1, \ldots, k_n)$\big) or \big($\M, v \models Y^n(k_1, \ldots, k_n)$ and $\M, v \not\models X^n(k_1, \ldots, k_n)$\big). Hence,  \big($\langle [k_1], \ldots, [k_n] \rangle \in v(X^n)$ and $\langle [k_1], \ldots, [k_n] \rangle \not\in v(Y^n)$\big) or \big($\langle [k_1], \ldots, [k_n] \rangle \in v(Y^n)$ and $\langle [k_1], \ldots, [k_n]\rangle\not \in v(X^n)$\big). Therefore, $v(X) \not= v(Y)$. Hence, $\M, v \not\models X=Y$.

\underline{$\chi$ is $\lambda X \psi(\iota Y \varphi)$.} (a) Assume that $\lambda X \psi(\iota Y \varphi)\in \Gamma$. By the witness property of $\Gamma \Rightarrow\Delta$, there exists a relational constant $K$ such that $\varphi^Y_K,\psi^X_K \in \Gamma$. By the induction hypothesis, $\M,v \models \varphi^Y_K$ and $\M,v \models\psi^X_K$. 
Suppose that for some $O_1,O_2\in G$, $\varphi^Y_{O_1},\varphi^Y_{O_2}\notin\Delta$ and $O_1=O_2\notin\Gamma$. 
By Lemma \ref{Prop41}, there are sequents $\Theta \Rightarrow \Lambda,\Pi\Rightarrow\Sigma,\Upsilon\Rightarrow\Phi\sqsubseteq\Gamma \Rightarrow\Delta$ such that $\mathcal{S} \vdash^{cf}_{\RLL}\Theta \Rightarrow \Lambda,\varphi^Y_{O_1}$, $\mathcal{S} \vdash^{cf}_{\RLL} \Pi\Rightarrow\Sigma,\varphi^Y_{O_2}$, and $\mathcal{S} \vdash^{cf}_{\RLL} O_1=O_2,\Upsilon\Rightarrow\Phi$. By weakening and the rule ($\iota^2_2\Rightarrow$), $\mathcal{S} \vdash^{cf}_{\RLL} \lambda X \psi(\iota Y \varphi),\Theta,\Pi,\Upsilon\Rightarrow\Lambda,\Sigma,\Phi$. This contradicts the $\mathcal{S}$-consistency of $\Gamma \Rightarrow\Delta$. Therefore, for all $O_1,O_2\in G$, $\varphi^Y_{O_1}\in\Delta$, or $\varphi^Y_{O_2}\in\Delta$, or $O_1=O_2\in\Gamma$. Hence, by the induction hypothesis, either, for all $O_1,O_2\in G$, $\M,v\not\models\varphi^Y_{O_1} $, or $\M,v\not\models\varphi^Y_{O_2} $, or $\M,v\models O_1=O_2$. Consequently, for all $O_1,O_2\in G$, if $\M,v\models\varphi^Y_{O_1} $  and $\M,v\models\varphi^Y_{O_2} $, then $\M,v\models O_1=O_2$. Thus, for any $ Y $-variant $ v' $ of $ v_K^X $, if $ \M, v' \models \varphi $, then $ v'(Y) = K $. Since $\M,v \models \varphi^Y_K$ and $\M,v \models\psi^X_K$, we obtain $ \M, v_K^X \models \psi $ and $ \M, v_K^X \models \varphi^Y_X $. So there is $K\in G$ such that $ \M, v_K^X \models \psi $, $ \M, v_K^X \models \varphi^Y_X $, and for any $ Y $-variant $ v' $ of $ v_K^X $, if $ \M, v' \models \varphi $, then $ v'(Y) = K $. 
Hence, $\M,v \models \lambda X \psi(\iota Y \varphi)$.

(b) Assume that $\lambda X \psi(\iota Y \varphi) \in \Delta$. By the witness property of $\Gamma \Rightarrow\Delta$, for each each relational parameter $B$, $\varphi^Y_B \in \Delta$, or $\psi^X_B \in \Delta$, or for some relational constant $K$, $\big(K=B \in \Delta$ and $\varphi^Y_K \in \Gamma\big)$. 
By the induction hypothesis, for each relational parameter $B$, 
$\M,v\not\models\varphi^Y_B$, or $\M,v\not\models\psi^X_B$, or for some relational constant $K$, $\big(\M,v\not\models K=B$ and $\M,v\models\varphi^Y_K\big)$. Therefore, $\M,v \not\models \lambda X \psi(\iota Y \varphi)$.
    
The other cases are treated similarly. 

Then one needs to show that $\M$ is a model of $\mathcal{S}$ but not of $H$. This can be done by the same technique as in \cite[Theorem 48]{LahavAvron}.
\end{proof}

\begin{theorem}
    For every sequent $\Gamma\Rightarrow\Delta$, $\vdash_{\RLL} \Gamma\Rightarrow\Delta$ implies that there exists a cut-free derivation of $\Gamma\Rightarrow\Delta$ in $\RLL$.
\end{theorem}
\begin{proof}
    Follows from Theorem \ref{Compl}.
\end{proof}

\section{Conclusion. Subjects for future research}\label{Concl}
The most apparent avenue for further research is the development of second-order variants of other theories of definite descriptions. This approach could be further generalized to examine higher-order theories of definite descriptions. Another route is to consider the modifications of Russellian theory of DD with a non-classical foundation. For example, in \cite{PetrukhinNelsonDD}, a version of Russellian theory based on Nelson's paraconsistent logic was introduced. One could try to develop its second- or higher-order version. Rather than adopting alternative theories, one can consider remaining within $ \RLL $ and conducting additional investigation:  for example, one could try to find a constructive proof of cut admissibility for this logic, building upon the existing proof for $ \RL $ as presented in \cite{IKRL}.

\paragraph{Acknowledgments.}
I would like to express my sincere gratitude and appreciation to Olena Dubchak for the everlasting inspiration. 
I am grateful to the anonymous reviewers for their insightful and constructive comments, which helped to improve the  presentation of the paper.  
Funded by the European Union (ERC, ExtenDD, project number: 101054714). Views
and opinions expressed are however those of the author(s) only and do not necessarily
reflect those of the European Union or the European Research Council. Neither the
European Union nor the granting authority can be held responsible for them.

\end{document}